\documentstyle[12pt]{article}
\newlength{\extraspace}
\newlength{\extraspaces}
\newcommand{\be}{\begin{equation}}
\newcommand{\ee}{\end{equation}}
\newcommand{\bea}{\begin{eqnarray}}

\newcommand{\eea}{\end{eqnarray}}

\newcommand{\nk}{\noindent}

\def\lsim{\mathrel{\rlap {\raise.5ex\hbox{$ < $}}
{\lower.5ex\hbox{$\sim$}}}}

\def\gappeq{\mathrel{\rlap {\raise.5ex\hbox{$>$}}
{\lower.5ex\hbox{$\sim$}}}}

\def\lappeq{\mathrel{\rlap{\raise.5ex\hbox{$<$}}
{\lower.5ex\hbox{$\sim$}}}}

\begin{document}

\begin{titlepage}

\begin{flushright}
ACT-02/97 \\
CERN-TH/96-353 \\
CTP-TAMU-04/97 \\
OUTP-97-04P \\
hep-th/9701144 \\
\end{flushright}
\begin{centering}
\vspace{.1in}
{\large {\bf On the Space-Time Uncertainty Relations of\\
Liouville Strings and $D$ Branes}} \\
\vspace{.2in}
{\bf G. Amelino-Camelia$^{a}$},
{\bf John Ellis$^{b}$},
{\bf N.E. Mavromatos$^{a,\diamond}$} \\
and \\
{\bf D.V. Nanopoulos}$^{c,d}$ \\
\vspace{.03in}
\vspace{.1in}
{\bf Abstract} \\
\vspace{.05in}
\end{centering}
{\small

Within a Liouville approach to non-critical string
theory, we argue for a non-trivial commutation
relation between space and time observables, leading
to a non-zero space-time uncertainty relation
$\delta x \delta t > 0$, which vanishes in the limit of 
weak string coupling.}
\vspace{0.2in}
\nk $^{a}$ University of Oxford, Dept. of Physics
(Theoretical Physics),
1 Keble Road, Oxford OX1 3NP, United Kingdom,   \\
$^{b}$ Theory Division, CERN, CH-1211 Geneva 23, Switzerland,  \\
$^{c}$ Center for
Theoretical Physics, Dept. of Physics,
Texas A \& M University, College Station, TX 77843-4242, USA\\
$^{d}$ Astroparticle Physics Group, Houston
Advanced Research Center (HARC), The Mitchell Campus,
Woodlands, TX 77381, USA. \\
\vspace{1.2in}
\begin{flushleft}
January 1997 \\
$^{\diamond}$ P.P.A.R.C. Advanced Fellow \\
\end{flushleft}
\end{titlepage}

\newpage
\baselineskip 12pt plus .5pt minus .5pt
\pagenumbering{arabic}
\pagestyle{plain} 

One important aspect of
recent developments~\cite{dbrane} in the
understanding of non-perturbative structures
in string theory is
renewed interest in the study of
modified uncertainty relations.
The conventional {\it critical string theory}
modification of
Heisenberg's uncertainty principle~\cite{heis}
was codified by
the {\it Enlarged Uncertainty Principle}~\cite{venezkonish}
\begin{eqnarray}
 \delta x \, \delta p \!\!& \ge &\!\! 1
+ {L_s^2} \, \delta p^2
~, \label{veneup}
\end{eqnarray}
where $L_s$ is the string length, {\it i.e.}, the square
root of the Regge slope: $L_s = \sqrt{\alpha '}$,
which leads to an absolute lower bound
on the measurability of distance:
\begin{eqnarray}
min \left[ \delta L \right] \!\!& = &\!\! L_s
~. \label{qgbounda}
\end{eqnarray}
Similar phase-space uncertainty relations
and measurability bounds for distances are believed to hold in
any theory involving a minimum length. 
In particular, an equation of the form of (\ref{qgbounda}), but
with $L_s$ replaced by the
Planck length $L_P$, has been proposed in the context of
some non-stringy approaches to quantum gravity\cite{padma}.
In addition, there has been discussion~\cite{yoneya} 
in the string-theory literature
of a possible non-trivial space-time uncertainty relation
of the type
\be
 \delta x \delta t \ge L_s^2
~.\label{dxdtls}
\ee
Both (\ref{veneup}) and (\ref{dxdtls})
have been reanalyzed in recent 
studies~\cite{dbrscatt,kogwheat,lizzi,yoneyanew}
of the non-perturbative solitonic structures in string theory known as
D~branes~\cite{dbrane}.
In particular, evidence has been found~\cite{dbrscatt}
in support of the idea that ``D~particles" (Dirichlet 0~branes)
could probe the structure of space-time down
to scales shorter than the string length,
raising the possibility that (\ref{veneup}) might have to be modified
in the D-brane context.

In parallel with these developments in the
literature on critical string theories,
there have recently been studies of novel uncertainty relations
and measurability bounds in the context of models of
quantum gravity theories which include intrinsic
microscopic mechanisms for quantum decoherence.
The emergent general expectation is that
decoherence effects should cause the
uncertainties involved in a measurement procedure
to grow with the time
needed for the measurement.
In particular, as discussed in ref.\cite{gacmpla}, 
the fact
that gravitational effects prevent one from relying
on the availability of {\it classical}
agents for the measurement procedure,
since the limit of infinite mass
leads to inconsistencies associated with the formation
of horizons,
leads to the following bound\footnote{Related work is reported 
in ref.~\cite{karo}.}
for the measurability of a distance $L$:
\begin{eqnarray}
min \left[ \delta L \right] \sim
\sqrt{ {T L_P^2 \over s}} \sim
\sqrt{ {L L_P^2 \over s}}
~,
\label{qgboundgac}
\end{eqnarray}
where $s$ is a length scale characterizing
the spatial extension of the devices ({\it e.g.}, clocks) 
used in the measurement, $T$ is the 
time needed to complete the procedure of measuring $L$,
and on the right-hand side we have taken into account
the fact that $T$ is typically proportional~\cite{gacmpla}
to $L$.
The bound (\ref{qgboundgac})
is always larger than $L_P$ for acceptable
values~\cite{gacmpla} of $s$, {\it i.e.} $L_P \! 
\lower .7ex\hbox{$\;\stackrel{\textstyle <}{\sim}\;$}
 \! s \! 
\lower .7ex\hbox{$\;\stackrel{\textstyle <}{\sim}\;$} \! L$,
and is maximal in the idealized scenario
$s \! \sim \! L_P$, in which case
\begin{eqnarray}
min \left[ \delta L \right] \!\!& = &\!\! 
\sqrt{ {L L_P}}
~.
\label{qgboundgacmax}
\end{eqnarray}
A candidate modified space-momentum uncertainty relation that 
leads to the bound (\ref{qgboundgac}) 
is given by~\cite{gacmpla,gacxt}
\begin{eqnarray}
 \delta x \, \delta p \!\!& \ge &\!\! 1 + {L_P^2} \, 
{T \over s} \, \delta p^2
~, \label{modup}
\end{eqnarray}
but it has recently become clear~\cite{gacxt} that the space-time
uncertainty relations 
\begin{eqnarray}
\delta x \, \delta t \ge {x {L_P^2 \over s}} 
 \label{newonegeneral}
\end{eqnarray}
or
\begin{eqnarray}
\delta x \, \delta t \ge t {L_P^2 \over s}
 \label{newonegeneralb}
\end{eqnarray}
would lead to the same bound.

The analysis of relations of the type (\ref{qgboundgac})-(\ref{newonegeneral}) 
has often remained at a rather heuristic level.
However, it has recently been realized~\cite{aemn} that Liouville 
(non-critical) string theories~\cite{emn}, in
which 
the target time is identified with
the Liouville mode~\cite{emn,kogan1}, 
and there is an intrinsic
microscopic mechanism for decoherence, provide a natural 
framework for the investigation of
such relations. 
Evidence in support of the validity of 
a measurability bound (\ref{qgboundgacmax})
in Liouville Strings has been provided in ref.~\cite{aemn}.
The analysis relied on the observation that the propagation 
of massless probes in the stochastic quantum-gravitational 
environment of Liouville stringy modes leads to a modification
of the dispersion relation of the massless probe.
In the case of a scalar-field prototype,
the resulting equation is a deformation of the Klein-Gordon equation,
with energy-dependent speeds for massless particles, and
this deformation leads to the bound (\ref{qgboundgacmax})
for the measurement of distances using such massless probes. 

In this paper, we present some new evidence for a possible
non-trivial space-time uncertainty relation, based on the
formulation of the sum over genera in Liouville
string theory, which entails the quantization of theory space,
as discussed in~\cite{emninfl,emnd}. As we
discuss in more detail below, this induces non-trivial
quantum behaviour in the time variable, which we identify
with the Liouville field.
Within the context of $D$-brane dynamics, in
which there is a departure from criticality related 
to the $D$-brane velocity~\cite{diffusion}, this leads
to a non-trivial space-time commutation relation 
of the form (\ref{dxdtls}). We find, however, that
this depends on the string coupling.

The canonical quantization of string theory space has
been discussed in~\cite{emninfl,emnd}. Motion in
this space is characterized as a renormalization-group flow in
the space of effective $\sigma$-model couplings $g^i$ on the
lowest-genus world sheet, with the flow variable identified
in turn with the Liouville field and the target time variable.
This flow is classical in the absence of higher-genus effects.
However, in their presence, the renormalization-group flow has been
shown to obey the relevant Helmholtz conditions which are
necessary and sufficient for the
canonical quantization of the $\sigma$-model couplings $g^i$:
\be
     [g^i, p_l ]=i \delta^{i}_l \, : \, p_l \equiv G_{lj}{\dot g}^j
~, \label{hel}
\ee
with an effective action related to an appropriate version of the
Zamolodchikov ${\cal C}$~function,
where $G_{ij}=<V_iV_j>$ is the Zamolodchikov metric in theory space, 
and the dot denotes differentiation with respect to the Liouville mode.

This summation over genera and the ensuing quantization clearly entail a 
modification of the conventional conception of space-time coordinates, though 
the exact nature of the resulting picture is still to be understood. Some 
aspects are, however, already apparent. The normalization of the Liouville 
field $\phi$ is such that the target time $t$ is given by~\cite{emn,aben}  
\be
t=Q[g^i] \, \phi_0 \qquad Q[g^i]^2 \equiv \frac{1}{3}({\cal C}[g^i]-25)
\label{target}
\ee
where $\phi_0$ is the world-sheet zero mode of $\phi$, and ${\cal C}[g^i]$ the
Zamolodchikov ${\cal C}$-function. In the presence of the sum over genera, the 
quantity $Q[g^i]$ becomes a $q$~number, as a result of the quantization of 
the $\sigma$-model couplings, and hence $t$ also becomes a quantum operator. 

Concerning the target-space coordinates, we observe that their tree-level
description as zero modes of the $\sigma$-model fields $X$
implies in non-critical strings an interdependence between these
coordinates and the background $\{ g^i \}$. For example,
in the case of the bosonic string that we consider here for simplicity,
for tachyonic backgrounds the following relation holds at tree-level 
\be
 < G_{lj}{\dot g}^j > = < [e^{i k X}] >
\label{tachy}
\ee
where $[\dots ]=1 + O(g)$ denotes a normal-ordered (renormalized)
product, which depends in general on the background $\{ g^i \}$. 
Eq.~(\ref{tachy}), which has trivial content 
in critical string theory, due to the on-shell 
conformal-invariance conditions $\beta^i \! = \! 0$, 
implies - upon appropriate inversion~\footnote{In order to 
establish this relation,
one would have to separate out the world-sheet zero mode in
$X=y+\xi(z,{\bar z})$,
and perform the path integration over $X$
involved in $< \dots >$ by first integrating over the quantum fluctuations
$\xi$, using the background-field method. 
The resulting relation between zero-mode integrals
also entails a relation between the respective 
integrands, which however can only be established up to the usual
kernel ambiguities.} - a non-trivial relation 
between the zero modes of the target space coordinates
and the backgrounds $G_{lj}{\dot g}^j$.
We expect that
upon summation over genera, which implies an appropriate path integration
over the $g^i$~\cite{emninfl,emnd},
this $c$-number relation would be turned into a (modified)
relation between the operator $G_{lj}{\dot g}^j$ and
the target-space coordinates.
Although the precise 
nature of the (correspondingly) quantized target-space
coordinates remains to be fully understood,
this suggests that the eqs.~(\ref{hel}) and (\ref{target})
may lead to non-trivial space-time commutation and uncertainty relations. 

The emergence of non-trivial space-time
commutation and uncertainty relations can be seen somewhat more directly
in the dynamical context of D branes treated as non-relativistic heavy
objects. In a $\sigma$-model formalism, the collective center of mass 
spatial coordinates of these soliton-like string backgrounds,
as well as the associated momentum coordinates,
can be described~\cite{dbrane,kogwheat,emnd,lizzi}
by couplings of relevant $\sigma$-model deformations.
Specifically, in
describing $D$-brane recoil induced by string-matter scattering
in a bosonic $\sigma$-model framework~\cite{kogwheat},
one introduces the two deformations
\bea
\int_{\partial \Sigma} y_i C(z) 
= y_i\int _{\partial \Sigma} d\tau \epsilon \Theta _\epsilon (X^0) 
\partial_n X^i
\label{recoila}\\
\qquad \int _{\partial \Sigma} u_{R,i} D(z) 
= u_{R,i}\int _{\partial \Sigma} d\tau \epsilon X^0 \Theta _\epsilon(X^0)
\partial_n X^i
\label{recoilb}
\eea
and the D-brane spatial coordinates are identified with the couplings $y^i$,
whereas the couplings $u_{R,i}$ correspond 
to the `renormalized'~\cite{diffusion,lizzi}
Galilean recoil velocity of the $D$ brane~\cite{kogwheat,diffusion,lizzi}. 
The D-brane mass is given~\cite{dbrane} by the
inverse of the string coupling $g_R$,
and therefore
its momentum coordinates are 
given by $p_i \! = \! u_{R,i}/g_R$. 
In (\ref{recoila}, \ref{recoilb}), the $X^i, i=1,2, \dots $
are understood to satisfy fixed Dirichlet boundary conditions
on the boundary $\partial \Sigma$, while for $X^0$ standard Neumann
boundary conditions are maintained. Also, in (\ref{recoila},
\ref{recoilb}), the quantity
$\epsilon$ is an infrared world-sheet scale related to the 
world-sheet size $L$ and the world-sheet ultraviolet 
cut-off $a$ by $\epsilon \equiv ({\rm ln}|L/a|)^{-1/2}$.

Since the phase-space coordinates
of the D brane
are themselves described 
by $\sigma$-model couplings,
after summation over genera the Liouville time\footnote{Note that,
in the D-brane case,
the zero mode $\phi_0$ of the Liouville field 
can be identified~\cite{diffusion}
with $\epsilon^{-1/2}$.}
is given by a $q$-number function of the generic form
\be
t = Q[y^i,p^i,g^i] \, \phi_0 
~,
\label{targetbrane}
\ee
where we have reserved the notation $g^i$ 
for couplings that are not related to the D-brane kinematics.
Moreover, the analysis in ref.~\cite{lizzi} indicates that 
the D-brane coordinates do not commute with their associated momenta.
This, together with (\ref{targetbrane}),
suggests the appearance of non-trivial commutation relations
between the space and time variables,
and associated space-time uncertainty 
relations. The determination of
the exact form of these uncertainty relations still requires substantial 
technical developments, along the lines 
set out in refs.~\cite{kogwheat,emnd,lizzi},
in order to understand better the structure 
of the function $Q[y^i,p^i,g^i]$ and the structure of the
commutation relations between
the D-brane coordinates and momenta.
At present, the best estimate of $Q$ comes from
the analysis
of Liouville dressing in the context of the non-critical
(if $\epsilon \ne 0$) 
$\sigma$-models describing $D$-brane recoil
induced by string-matter scattering.
The recoil operators $C$ and $D$,
involved in (\ref{recoila}) and (\ref{recoilb}) 
respectively,
are known~\cite{kogwheat} 
to have a small  negative  
anomalous dimension $ - \frac{1}{2}\epsilon ^2$.
This non-criticality leads to Liouville 
dressing, and 
via an application of Zamolodchikov's ${\cal C}$-theorem 
one finds in this $D$-brane case that~\cite{diffusion,lizzi}
\be 
Q^2 \sim \int_{\phi^*}^\phi d\phi' \frac{\partial}{\partial \phi'}
{\cal C}[\phi '] =
-\int_{\phi_*}^\phi d\phi ' \beta^C G_{CC} \beta^C =
-\int_{\phi_*}^\phi \frac{d\phi '}{(\phi ')^2} \frac{u_R^2}{g_R} 
\sim \frac{u_R^2}{g_R\phi} 
\label{qphi}
\ee
where $\phi^*$ is an infrared fixed point on the world-sheet,
and we have used the results of ref. \cite{kogwheat} to express 
the two-point function of the $C$ recoil operator
in terms of the local renormalization-group scale 
on the world sheet (Liouville mode): 
\be
G_{CC} \sim 2|z|^4<C(z)C(0)> \sim 
\frac{1}{g_R \phi^2}
\label{cc}
\ee
Using the relations (\ref{targetbrane}) and (\ref{qphi}),
we can obtain some preliminary evidence on the structure
of the commutation relations for the Liouville D-brane
space-time coordinates.
Considering for simplicity the 
case of a 1+1-dimensional space-time,
it follows from (\ref{targetbrane}) and (\ref{qphi}) that
\be 
[y, t] \sim [y, p] \, \sqrt{g_R \phi} L_s^2
~.\label{nontrivial}
\ee
We use the preliminary (lowest-order)
results of ref.~\cite{lizzi} for the phase-space coordinates of 
the D-brane, which indicate the following form of
commutation relation 
\be 
    [y^i, p_j] =i\delta^i _j + {\rm stringy~corrections}
~,\label{uncert}
\ee
{\it i.e.}, the space and momentum coordinates
of the D brane form, as a first approximation, an ordinary 
quantum-mechanical phase space.

Combining (\ref{nontrivial}) and (\ref{uncert}),
one finally obtains the non-trivial space-time
commutation relation
\be
   [y, t] \sim i \sqrt{g_R \phi} L_s^2
~.\label{dbranunc}
\ee
In order to evaluate the
corresponding space-time uncertainty relation, 
by analogy with the uncertainty relations characterizing
ordinary quantum mechanics, we need to estimate the
minimum value of the
expectation value $<\sqrt{g_R \phi} >$ in a physical state.
At present, the space of physical string states, and especially 
its measure, is not well understood. However, if one interprets a
world-sheet $\sigma$-model 
partition function as a 
target-space string wave function, 
the quantity $< \sqrt{g_R \phi} >$ 
could be evaluated as a
$\sigma$-model vacuum expectation value.
In the Liouville approach discussed here,
such $\sigma$-model vevs include the
summation over world-sheet topologies, which represents
a path integration over stringy backgrounds~\cite{emninfl,emnd},
as well as the integration over the $\sigma$-model fields $X^i$ and
$\phi$. In the string-theory language, such vevs are string field-theory 
vevs. With this interpretation in mind, we estimate
\be
 \delta y \delta t \, \ge \, < \sqrt{g_R \phi} > L_s^2
\, \sim \, \sqrt{g_R} L_s^2
~, \label{dxdtaemn}
\ee
where the approximation on the right-hand side
follows if we estimate
$< \sqrt{\phi} >$ to be of $O(1)$.

We observe that the uncertainty relation (\ref{dxdtaemn}) 
is $y$- and $t$-independent,
just like (\ref{dxdtls}).
However, we also note that modifications are to be expected from effects
beyond our leading-order analysis, because of non-trivial
contributions coming from terms
on the right-hand-side of eq.(\ref{uncert})
that we ignored.
Even at the present level of analysis,
there is the important difference of a factor of $\sqrt{g_R}$
between eq.~(\ref{dxdtaemn}) and eq.~(\ref{dxdtls}).
As a result, we find
no space-time uncertainty relation 
in the weak-coupling limit $g_R \rightarrow 0 $.
This difference between the uncertainty relations
characterizing the dynamics of the Liouville D branes~\cite{emnd,diffusion}
considered here and the corresponding relations 
of~\cite{dbrane,dbrscatt,yoneyanew}
originates from the structure of the relevant
recoil operators~\cite{kogwheat}
and the associated~\cite{lizzi}  
space-momentum uncertainty relation (\ref{uncert}).

It would be of great interest to understand more
deeply the group-theoretical structure of the non-trivial
space-time commutation relation (\ref{dbranunc})
and uncertainty relation (\ref{dxdtaemn})
that we have found above. Whilst one might expect
the underlying structure to be $O(d)$ invariant,
we see no reason to expect that Lorentz invariance
would be retained, since this is a property derived
from critical string theory. It is attractive to
speculate that some sort of quantum group structure,
such as $\kappa$-Poincar\'e~\cite{kpoin}, might be relevant, but
we have no formal evidence for this hypothesis.
A proper setting for the preliminary 
investigation of these issues might be provided by 
field theories based on such uncertainty relations,
in analogy with the analysis given in ref.~\cite{kempf}
of field theories with modified phase-space uncertainties.
However, such an analysis goes beyond the scope of this note.

Our essential point here has been to observe that as
soon as one broadens the scope of string theory to include
non-critical configurations, as is necessary to accommodate
fluctuations in the classical string background, e.g. the 
case of $D$-brane quantum recoil~\cite{emnd,diffusion},
time
becomes a $q$ number related to the Liouville field, and as
such acquires non-trivial commutation relations with spatial
coordinate observables, implying in turn non-trivial
space-time uncertainty relations.

\newpage

\baselineskip 12pt plus .5pt minus .5pt

\end{document}